\documentclass[useAMS,usenatbib,usegraphicx]{mn2e}

\title[An overdensity of massive YSOs around \emph{Spitzer} mid-IR bubbles]{The statistics of triggered star formation:\\  An overdensity of massive YSOs around \emph{Spitzer} bubbles}
\author[]{M. A. Thompson$^{1}$\thanks{E-mail: m.a.thompson@herts.ac.uk}, J. S. Urquhart$^{2,3}$, T. J. T. Moore$^{4}$ and L. K. Morgan$^{4}$\\
$^{1}$Centre for Astrophysics Research, Science \& Technology Research Institute,
University of Hertfordshire, College Lane,\\ Hatfield, Herts, AL10 9AB, UK\\
$^{2}$Australia Telescope National Facility, CSIRO Astronomy and Space Science, PO Box 76, Epping NSW 1710, Australia\\
$^{3}$ Max Planck Institut f\"ur Radioastronomie, Auf dem H\"ugel 69, 53121 Bonn, Germany\\
$^{4}$Astrophysics Research Institute, Liverpool John Moores University, Twelve Quays House, Egerton Wharf, Birkenhead, CH41 1LD, UK}
\begin{document}

\date{}

\pagerange{\pageref{firstpage}--\pageref{lastpage}} \pubyear{2011}

\maketitle

\label{firstpage}

\begin{abstract}
We present a detailed statistical study of massive star formation in the environment of 322 \emph{Spitzer} mid-infrared bubbles by using the RMS survey for massive Young Stellar Objects (YSOs). Using a combination of simple surface density plots and a more sophisticated angular cross-correlation function analysis we show that there is a statistically significant overdensity of RMS YSOs towards the bubbles. There is a clear peak in the surface density and angular cross-correlation function of YSOs projected against the rim of the bubbles. By investigating the autocorrelation function of the RMS YSOs we show that this is not due to intrinsic clustering of the RMS YSO sample. RMS YSOs and \emph{Spitzer} bubbles are essentially uncorrelated with each other beyond a normalised angular distance of two bubble radii. The bubbles associated with RMS YSOs tend to be both smaller and thinner than those that are not associated with YSOs. We interpret this tendency to be due to an age effect, with YSOs being preferentially found around smaller and younger bubbles.  We find no evidence to suggest that the YSOs associated with the bubbles are any more luminous than the rest of the RMS YSO population, which suggests that the triggering process does not produce a top heavy luminosity function or initial mass function. We suggest that it is likely that the YSOs were triggered by the expansion of the bubbles and estimate that the fraction of massive stars in the Milky Way formed by this process could be between  14 and 30\%.

\end{abstract}


\section{Introduction}
\label{sect:intro}

Massive stars drive large amounts of energy into the surrounding interstellar medium (ISM) via their winds and ionising radiation,  sculpting the ISM into a series of bubble and shell-like structures \citep{elmegreen2011}. One particular class of shells or bubbles are those surrounding expanding HII regions. The UV illumination excites emission from Polycyclic Aromatic Hydrocarbons (PAHs) at the bubble rims and also heats dust grains within the HII region. The former can be readily observed at 8\,$\mu$m by IRAC onboard the \emph{Spitzer} Space telescope \citep{churchwell2006,churchwell2007} and the latter at longer wavelengths using e.g.~MIPS on \emph{Spitzer} \citep{churchwell2007,watson2008} or PACS onboard the Herschel Space Observatory \citep{zavagno2010b,zavagno2010a}. 

The expansion of these HII regions  is of extreme interest to studies of star formation as their expansion may trigger new generations of star formation into being within the molecular material surrounding  the bubbles. There are two major triggering mechanisms that have been put forward so far: Radiative Driven Implosion and the Collect \& Collapse process. In Radiative Driven Implosion \citep[RDI:][]{bertoldi1989,lefloch1994, miao2009, bisbas2009,bisbas2011}  the expanding ionisation front of the HII region drives a D-type shock into molecular clouds surrounding the HII region, triggering the collapse of sub-critical clumps within the clouds. Theoretical models of RDI can successfully explain the morphology of BRCs \citep{miao2009,bisbas2009,bisbas2011} and observations suggesting that star formation is concentrated along the central axis of the clouds \citep{sugitani1999,sugitani2000,bisbas2011}.

The Collect \& Collapse process \citep{elmegreen1977,whitworth1994,whitworth2002}, on the other hand, does not require the presence of pre-existing molecular structures. In this case the expansion of the HII region sweeps up the surrounding low density material into a shell surrounding the HII region. At a certain point (generally after a few Myr) this shell becomes self-gravitating, fragments and collapses to form dense molecular clumps that eventually collapse to form stars. The fragments formed in the Collect \& Collapse process tend to be massive \citep[a few hundred M$_{\odot}$,][]{whitworth1994} and so this process could naturally explain the hierarchical nature of massive stellar clusters \citep[e.g.][]{bastian2005,oey2005}. It should be noted that neither mechanism excludes the other as  a means of triggering star formation. Indeed, within a single HII region both RDI and Collect \& Collapse may operate together \citep{deharveng2005}. 

However, recent models of the ionising feedback from massive stars \citep{dale2011} suggest that neither mechanism is responsible and that the UV illumination from these stars simply erodes low density material rather than shaping its evolution and eventual collapse. In this scenario, the role of triggering in star formation is predicted to be minimal. \citet{walch2011} find  that the location of the dense clumps around the edge of HII regions (such as RCW 120) reflects the pre-existing cloud structure and their formation does not require the Collect \& Collapse process. However, rather than neglecting triggering as in \citet{dale2011}, they suggest that stars may form by global implosion of the pre-existing structures (Enhancement of initial Density substructure and simultaneous Global Implosion, or EDGI).

Observational studies of triggered star formation have so far mainly focused on photoionised globules \citep[Bright-Rimmed Clouds or BRCs,][]{sugitani1991,sugitani1994} found at the edges of optically visible HII regions \citep{thompson2004a, urquhart2004,morgan2004, urquhart2006,morgan2008} or on the rims of HII regions selected to show a relatively simple morphology \citep{deharveng2005,zavagno2006,pomares2009}. Recent studies of the latter have been able to take advantage of the large catalogue of infrared bubbles discovered in the GLIMPSE survey \citep{churchwell2006,churchwell2007}. The properties of these bubbles are consistent with expanding HII regions \citep{deharveng2010, watson2009}, although perhaps more consistent with a ring rather than bubble morphology \citep{beaumont2010}. 

However, one difficulty with many of these studies is that they are often phenomenological in nature, concentrating upon the visual identification of YSOs or protostars in regions where one might have a prior expectation that they may have been triggered \citep[e.g.][]{zavagno2006,zavagno2007,deharveng2010}. Such studies cannot attack the central problem in triggered star formation, which is to identify the origin of the discovered star formation (we refer to this as the ``Origin Problem''). When trying to identify star formation as being triggered one must first exclude the possibility that the star(s) would have formed spontaneously without the influence of the trigger. The Origin Problem is particularly intractable when considering individual objects --- without a good understanding of the initial conditions involved it is almost impossible to categorise an individual star-forming region as being triggered or spontaneous.

In this paper we attempt to move beyond the current phenomenological approach by carrying out a detailed statistical study of star formation around the 322 \emph{Spitzer} bubbles of \citet{churchwell2006}. While the Origin Problem is almost impossible to solve for individual star-forming regions, by considering  the global properties of a large sample it may be possible to infer the presence of triggering in a statistical sense. These studies are made easier by the existence of large,  uniformly selected and well-understood  star formation surveys such as the RMS survey \citep{urquhart2008}. The RMS survey has the goal of identifying every massive young stellar object (YSO) in the Milky Way and comprises an initial infrared selection followed by thorough multi-wavelength follow-up to rigorously classify each object and determine its physical properties \citep{urquhart2009b,urquhart2009a,mottram2011a,urquhart2011}.

\begin{figure}
\includegraphics[width=84mm,angle=0]{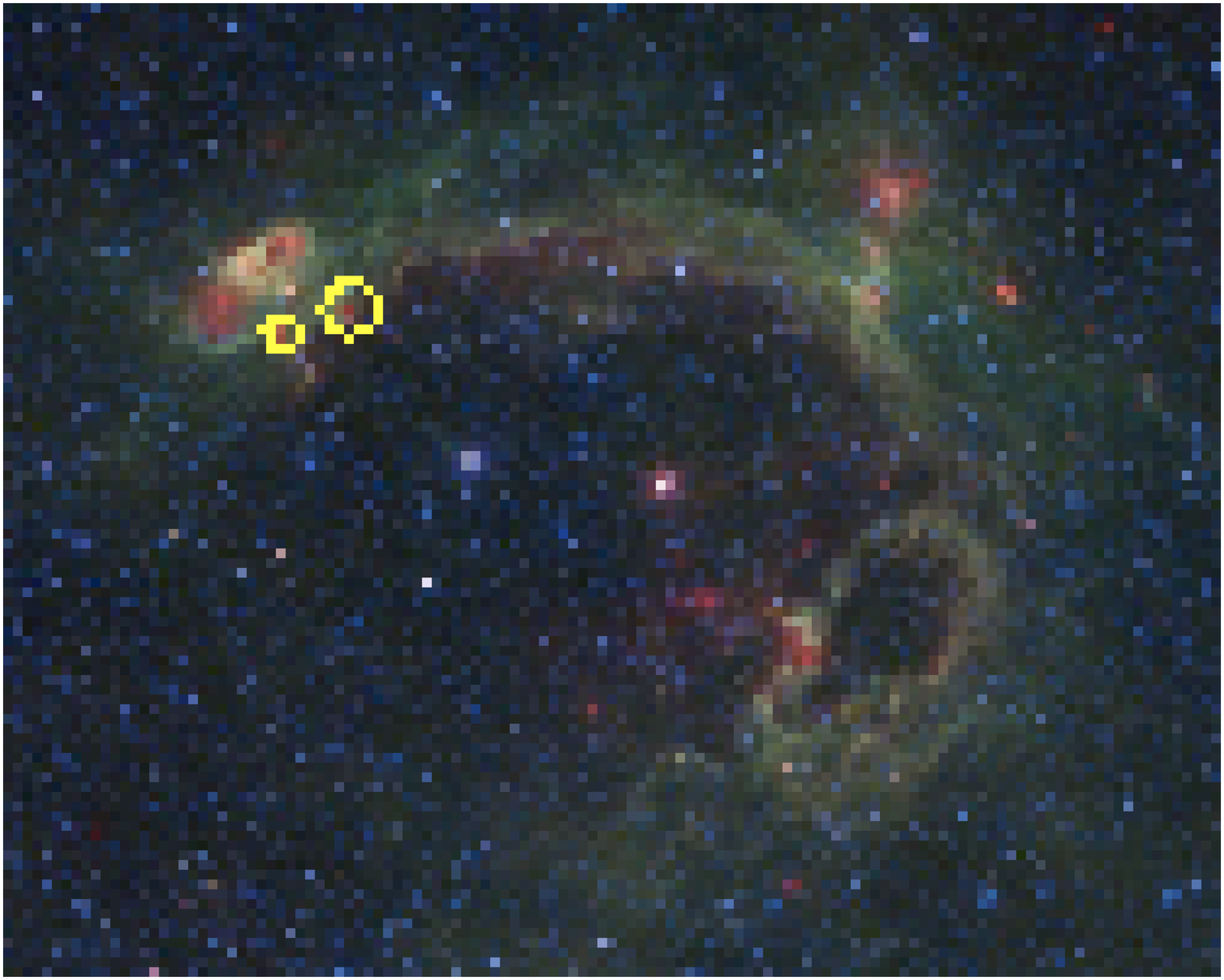}
\caption{Three colour GLIMPSE image of a  \emph{Spitzer} bubble (N109 from the \citealt{churchwell2006} catalogue), showing the strong extended 8 $\mu$m PAH and 24 $\mu$m dust emission tracing the bubble rim. The colour coding in the image is 24 $\mu$m (red),  8  $\mu$m (green) and 4.5 $\mu$m  (blue).  The 24 $\mu$m image is taken from the MIPSGAL survey \citep{carey2009}. N109 is one of the largest bubbles in the \citet{churchwell2006} catalogue, with a mean radius of 14.8\arcmin, and is also the site of numerous smaller bubbles. The positions of  two objects from the RMS YSO sample are indicated by green circles.}
\label{fig:bubble}
\end{figure}

The RMS survey covers a greater area than the GLIMPSE-I survey region in which the \citet{churchwell2006} bubbles have been identified and  is complete to YSOs of luminosity $\ge$10$^{4}$ L$_{\odot}$ out to the furthest bubble in the \citet{churchwell2006} catalogue. Hence we can conduct a survey for recent massive star formation around \emph{all} of the \emph{Spitzer} bubbles (see Fig.~\ref{fig:bubble} for an example).  Note that the RMS survey does not include Galactic latitudes $|l|\le10$\degr\ for reasons of cofusion and so we do not consider here the bubble catalogue of \citet{churchwell2007} which is based solely upon the GLIMPSE-II survey region (i.e.~$|l|\le10$\degr). On the other hand we must keep in mind the intrinsic subjective biases in the catalogue of \emph{Spitzer} bubbles, due to their manual by-eye  identification by a number of independent observers \citep{churchwell2006}. These biases are discussed in detail in Section 2 of \citet{churchwell2006}. The current catalogue is likely to be highly incomplete, particularly to small bubbles. \citet{churchwell2006} estimate a completeness on the order of $\sim$50\%, which is being borne out by early results from the citizen science Milky Way Project\footnote{\texttt{http://www.milkywayproject.org}} (Simpson et al.~2012, in prep).

Nevertheless, the \citet{churchwell2006} catalogue of bubbles currently represents the most complete and well-studied catalogue of HII regions with a simple morphology that lends well to statistical studies of their YSO distribution. In this paper we present such a study, paying particular attention to the angular distribution of YSOs around the bubbles and potential differences within the YSO population. Our aims are to provide statistical evidence that the star formation associated with the bubbles may have been triggered and to investigate the properties of the global population of bubbles and YSOs. In addition, the methods that we demonstrate in this paper will be readily applicable to the  larger and more complete Milky Way Project bubble sample when it becomes available.

The paper is structured as follows. In Sect.~\ref{sect:environment} we study the star forming environment of the \emph{Spitzer} bubbles using a simple surface density approach, followed by a more sophisticated analysis of the angular cross-correlation function of RMS YSOs and \emph{Spitzer} bubbles. Sect.~\ref{sect:discussion} combines the results of the star forming environment analysis with the observed properties of the bubbles and YSOs to demonstrate that star formation is clearly enhanced toward the bubbles. We speculate on the likely origin of this star formation and show that it is likely that the bubbles predate the formation of the RMS YSOs. Finally in Sect.~\ref{sect:conclusions} we present a summary of our conclusions and results.

\section{The star forming environment of \emph{Spitzer} bubbles}
\label{sect:environment}

\subsection{The surface density of YSOs}
\label{sect:surfdens}

As a first approach to studying  the distribution of RMS YSOs around \emph{Spitzer} bubbles we simply measured the number counts of RMS YSOs expressed as a function of angular separation from the bubble centres.  Our sample of RMS YSOs is comprised of the objects classified as either YSO or UC HII in the RMS database\footnote{http://www.ast.leeds.ac.uk/RMS}, see \citet{urquhart2008} for a description of the RMS database and its classification system.  YSO and UC HII sub-classifications both represent young, recently formed and predominantly massive stellar objects that enable us to trace the distribution of recent star formation around the bubbles. Hereafter we refer to this combined population as the RMS YSO sample. Within the area covered by the GLIMPSE I survey \citep{benjamin2003} there are  846 objects within the RMS YSO sample and 322 bubbles from the \citet{churchwell2006} catalogue.

\begin{figure}
\includegraphics[width=84mm,angle=-90]{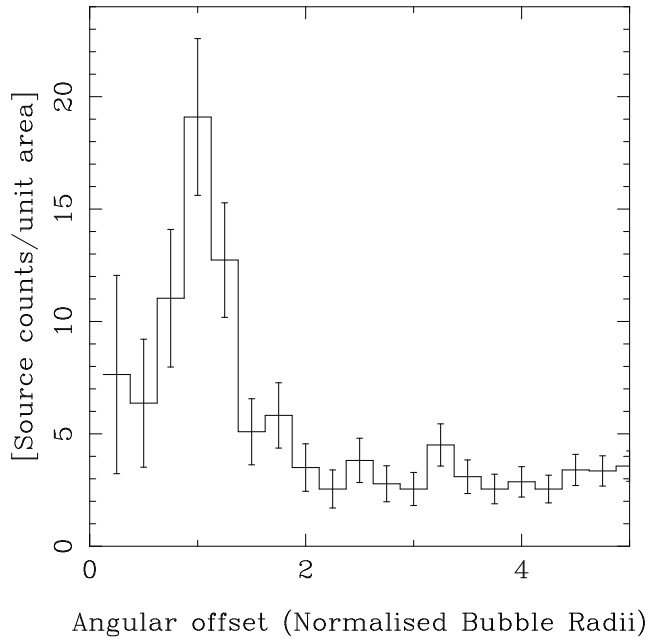}
\caption{Histogram of the number counts of RMS YSOs (comprising YSO and UC HII classifications from the RMS database) as a function of angular distance from the centre of \emph{Spitzer} bubbles. The distance is expressed in terms of normalised bubble radius. The number counts are scaled by the area of the annulus corresponding to each bin and thus represent a surface density of RMS YSOs. Error bars are determined via Poisson statistics.}
\label{fig:surfdens}
\end{figure}

 In order to account for the different angular radii of the bubbles we divided the angular separation of each RMS object from a particular bubble by the mean radius of the bubble ($\left<R\right>$, column 9 in the catalogue of \citealt{churchwell2006}), i.e.~expressing the angular separation in bubble radii rather than arcminutes. Each bin  represents an annulus around the centre of each bubble. The surface area of the annulus thus naturally increases with increasing radius and so to obtain the surface density of RMS YSOs we scale the counts in each bin by the surface area of each corresponding annulus.  A histogram of these scaled number counts is shown in Fig.~\ref{fig:surfdens}.

Fig.~\ref{fig:surfdens} shows a clear peak in the number of RMS YSOs at a separation of 1 normalised bubble radius. At greater angular radii the number of RMS YSOs falls sharply, reaching a constant background level of $\sim$ 3 sources by 2 bubble radii. Within an angular  radius of 2 normalised bubble radii the number of RMS YSOs is demonstrably higher than at angular radii greater than 2 bubble radii.  The surface density of YSOs projected against \emph{Spitzer} bubbles is thus higher than regions external to the bubbles, with a clear peak in the surface density projected against the the rims of the bubbles.

The \emph{Spitzer} bubbles are relatively elliptical, with typical eccentricities between 0.6 and 0.7. As we normalise by the mean bubble radius $\left<R\right>$ this will have the effect that we incorrectly calculate the true normalised radius of each RMS YSO from the bubble centre, potentially broadening the observed peak in surface density. The position angles of the elliptical fits to the bubbles are not listed in \citet{churchwell2006}, but these measurements were kindly made available by Matt Povich (Povich, priv.~comm.) so that we could examine the effect of using the true radius of the bubble instead of the mean radius. We found that there is no significant difference between scaling the distance of the RMS YSOs with the true bubble radius and the mean radius. This is more than likely due to the fact that the angular resolution of our histogram in Fig.~\ref{fig:surfdens} is limited to 0.25 bubble radii by the need to obtain sufficient RMS YSOs in each bin. At this resolution the worst-case error in radius (i.e.~between mean radius $\left<R\right>$ and the semi-major axis $a$) is slightly larger than the width of one bin in Fig.~\ref{fig:surfdens}. 

We further subdivided our RMS YSO sample into its constituent YSO and UC HII sub-samples to investigate trends in the separate distributions of YSOs and UC HIIs, for example in evolutionary status versus radius. We found that there is no significant difference between the two sub-samples, the histograms of separate YSO and UC HII sub-samples are indistinguishable from the combined RMS YSO sample. Again, this may be due to the limited sample statistics that we currently have, or this may indicate that gradients in evolutionary status around the bubbles are either not present or, if present, are  on smaller angular scales than resolved by our study.

\begin{figure}
\includegraphics[width=84mm]{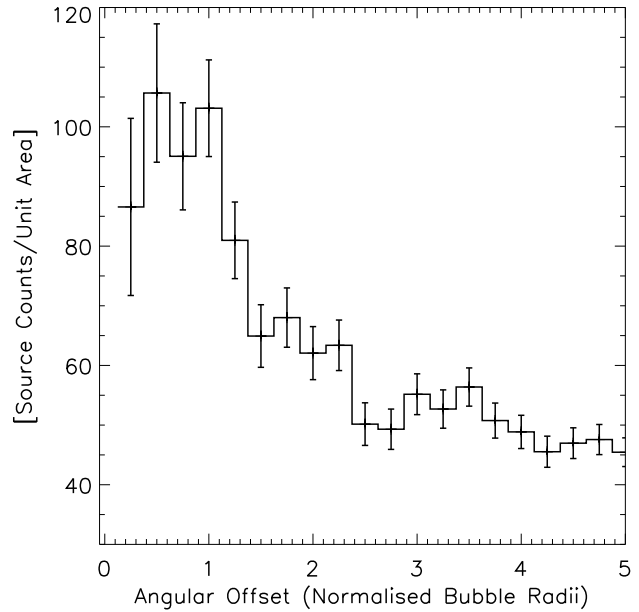}
\caption{Histogram of the number counts of \citet{robitaille2008} Intrinsically Red Objects  as a function of angular distance from the centre of \emph{Spitzer} bubbles. The distance is expressed in terms of normalised bubble radius. The number counts are scaled by the area of the annulus corresponding to each bin and thus represent a surface density. Error bars are determined via Poisson statistics.}
\label{fig:glimpse_surfdens}
\end{figure}

In order to confirm this result we also carried out the same analysis on the \cite{robitaille2008} catalogue of intrinsically red sources selected from the \emph{Spitzer} GLIMPSE survey. The Robitaille catalogue contains a much greater number of objects than the RMS survey, though at the expense of contamination by an uncertain fraction of AGB stars \citep{robitaille2008}. The surface density of objects from the Robitaille catalogue is shown in Figure \ref{fig:glimpse_surfdens} and displays a similar distribution to the RMS YSO sample, with a higher surface density  towards the bubbles that sharply drops off to a uniform background level. 

The background level is much higher than the RMS YSO sample, as would be expected due to the higher surface density of  \citet{robitaille2008} intrinsically red sources compared to the RMS catalogue \citep{urquhart2008}. The distribution of \citet{robitaille2008} objects does not peak at 1 bubble radius, but instead exhibits a relatively flat distribution out to 1 bubble radius. As the RMS catalogue is constrained to star forming objects at an early evolutionary state (YSOs and UC HII regions), whereas the \citet{robitaille2008} catalogue is not, this may indicate the presence of an evolutionary gradient across the bubbles. Further classification of the \citet{robitaille2008} sample and investigation of their star-forming nature are required to prove this hypothesis.

We must also explore the possibility that there may be an intrinsic bias in the distribution of both the RMS and \citet{robitaille2008} catalogues around the \emph{Spitzer} bubbles due to the common mid-infrared bands used to detect both the bubbles and RMS/Robitaille objects. Although
the bubbles are principally identified via their extended PAH emission at 8 $\mu$m and the RMS YSOs and Robitaille intrinsically red sources are predominantly point infrared sources, the complex mid-infrared environments of the bubbles may lead to a bias in the identification of point sources at their rims.  We investigate this possibility by examining the distribution of  6.7 GHz methanol masers drawn from the Methanol MultiBeam (MMB) Survey \citep{green2009} around the \emph{Spitzer} bubbles. 6.7 GHz methanol masers are thought to exclusively trace young sites of massive star formation \citep[e.g.][]{menten1991}, and thus allow us to trace the distribution of massive YSOs around the bubbles independently of their mid-infrared emission.

\begin{figure}
\includegraphics[width=84mm,angle=-90]{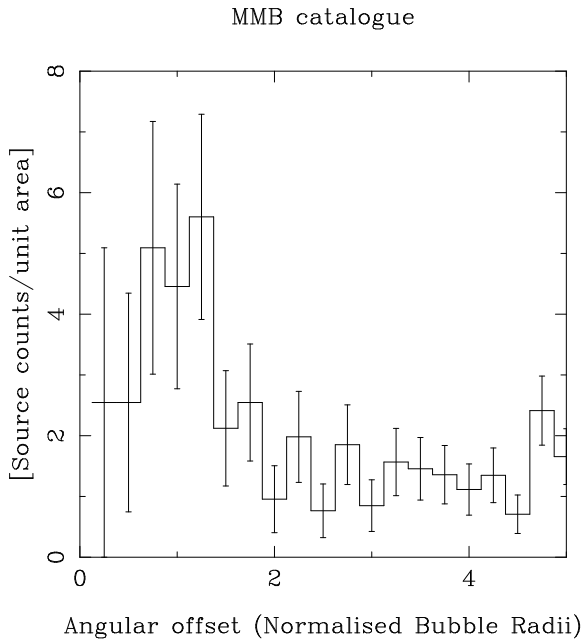}
\caption{Histogram of the number counts of MMB 6.7 GHz masers as a function of angular distance from the centre of \emph{Spitzer} bubbles. The distance is expressed in terms of normalised bubble radius. The number counts are scaled by the area of the annulus corresponding to each bin and thus represent a surface density. Error bars are determined via Poisson statistics.}
\label{fig:mmb_surfdens}
\end{figure}

The MMB survey currently occupies a longitude range between $l=186$ and $l=20$, i.e. excluding the  range $20\le l \le 186$, and so only the bubbles in the southern GLIMPSE survey region are presently covered by the MMB survey. The individual masers in the MMB catalogue have had their positions interferometrically  determined to sub-arcsecond precision and the maser detections are reported in \citet{caswell2009}, \citet{green2009}, \citet{caswell2010}, \cite{green2010}, \cite{caswell2011} and Green et al.~(2012, in press). We plot the surface density of  6.7 GHz MMB masers around the southern \emph{Spitzer} bubbles in Fig.~\ref{fig:mmb_surfdens}.

Fig.~\ref{fig:mmb_surfdens} displays a very similar distribution of masers to that of RMS YSOs and \cite{robitaille2008} intrinsically red sources, albeit with larger error bars due to the smaller sample size. There is a clearly distinguished peak in the maser distribution at an angular offset of 1 bubble radius and the surface density of 6.7 GHz masers drops to a roughly constant background level beyond an offset of 2 bubble radii. The peak in the surface density of 6.7 GHz masers appears to be broader than the corresponding peak in the surface density of RMS YSOs, however the lower signal-to-noise of the MMB surface density makes it difficult to interpret this difference as a real effect.

All three independently selected YSO catalogues (RMS, \cite{robitaille2008} red sources and MMB 6.7 GHz masers) display very similar surface density distributions and we thus conclude that the increase in surface density of YSOs towards the bubble rims is a real effect.  Given this similar behaviour between catalogues, and the currently more comprehensive knowledge of the properties of the  RMS YSOs \citep[e.g.][]{urquhart2011,mottram2011a,urquhart2009b,urquhart2009a}, 
we restrict our further analysis to the RMS YSO catalogue. Although the \citet{robitaille2008} red source catalogue has greater sample statistics and is likely to be predominantly comprised of YSOs there is a much greater likelihood of contamination by AGB stars and other non-YSO types than in the RMS YSO catalogue. Similarly, the lower numbers of objects in the MMB 6.7 GHz maser catalogue favours the continuation of our study using the larger and much more studied RMS YSO sample.

\subsection{The angular cross-correlation of bubbles and YSOs}
\label{sect:crosscol}

As a refinement of our simple surface density approach  we also investigated the distribution of RMS YSOs around the bubbles  using an angular two-point cross-correlation analysis, a technique more commonly used to determine the clustering properties of galaxies  \citep[e.g.][]{smith1995,ghirlanda2006,wang2011,bradshaw2011}. The correlation function defines the probability of finding a population of objects at a particular angular separation from a different second population. Here, we used the RMS YSO sample  as our first population ($D_{1}$) and the \cite{churchwell2006} catalogue of infrared bubbles for our second population ($D_{2}$). We calculated the angular cross-correlation using the estimator of \cite{landy1993}, modified for the cross-correlation between population 1 and 2 using the equation of \cite{bradshaw2011}, i.e.

\begin{equation}
\omega(\theta) = \frac{N_{D_{1}D_{2}} - N_{D_{1}R_{2}} - N_{R_{1}D_{2}} + N_{R_{1}R_{2}}}{N_{R_{1}R_{2}}}
\end{equation}

where $N_{D_{1}D_{2}}$ represents the normalised number counts at an angular separation of $\theta$ of RMS source-bubble pairs, $N_{D_{1}R_{2}}$ and $N_{R_{1}D_{2}}$ the counts of real and random catalogues of RMS source-bubble pairs (and vice-versa), and $N_{R_{1}R_{2}}$ the counts between two random catalogues of RMS objects and Spitzer bubbles. As in Sect.~\ref{sect:surfdens} we scaled $\theta$ to the radius of the individual Spitzer bubble in each pair. To avoid introducing high levels of noise through the randomly generated catalogues we performed 50 realisations of each catalogue, taking the mean of the results to determine $\omega(\theta)$. The errors  on $\omega(\theta)$ were calculated by a bootstrapping approach \citep[e.g.][]{ghirlanda2006,bradshaw2011}. The Spitzer bubble catalogue was divided into 100 randomly chosen bootstrap catalogues  each matching the original catalogue size (with replacement). The angular cross-correlation was determined for each bootstrap catalogue and the resulting 1$\sigma$ error in $\omega(\theta)$ is given by the standard deviation in $\omega(\theta)$ from the 100 random bootstrap samples.  

\begin{figure}
\includegraphics[width=84mm,angle=-90]{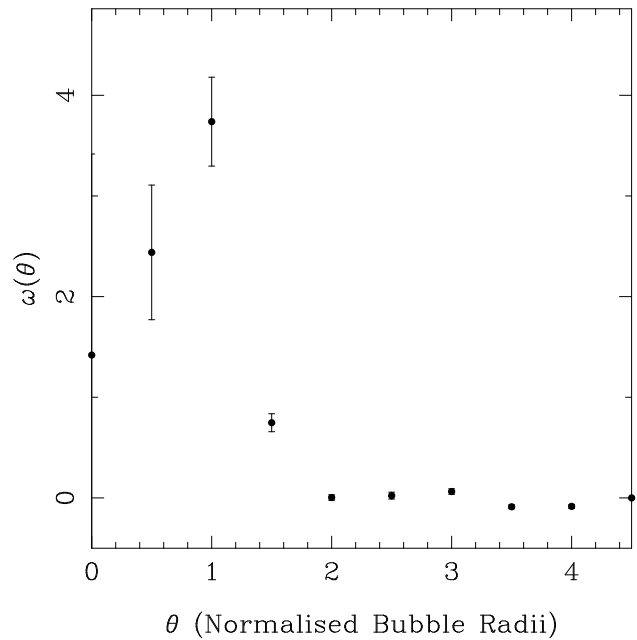}
\caption{The angular cross-correlation of the RMS YSO sample and the catalogue of \emph{Spitzer} bubbles as a function of normalised bubble radius. }
\label{fig:crosscol}
\end{figure}

The resulting angular cross-correlation function is plotted in Figure \ref{fig:crosscol}, which reveals almost exactly the same distribution as seen in the surface density distribution shown in Figures \ref{fig:surfdens}--\ref{fig:mmb_surfdens}.  The RMS YSO sample is found to be strongly correlated with the \emph{Spitzer} bubble catalogue, particularly at the radius corresponding to the rim of the bubbles where the correlation peaks. This peak is significant at the 9$\sigma$ level. The cross-correlation drops sharply beyond the peak at 1 bubble radius and beyond a distance of 2 bubble radii the cross correlation decreases to essentially zero.  This indicates that the probability of finding an RMS YSO near a \emph{Spitzer} bubble is markedly greater at an angular radius of 1 bubble radii, and that  beyond an angular distance of 2 bubble radii the RMS YSO population are essentially uncorrelated with the presence of a \emph{Spitzer} bubble.

\subsection{The angular autocorrelation of RMS YSOs}
\label{sect:autocol}

Finally we investigate the clustering within the RMS YSO sample, in order to determine whether the previous results in Sect.~\ref{sect:surfdens} and \ref{sect:crosscol} are simply due to an intrinsic angular clustering scale within the RMS catalogue that happens to correspond to the typical angular size of a \emph{Spitzer} bubble. The median bubble radius of all the  bubbles listed in the \citet{churchwell2006} catalogue is 1.1 arcmin, with a large observed range stretching between 0.14 arcmin and 14 arcmin. However, few bubbles possess extreme radii and two thirds of the sample have radii between 1--3 arcmin. If the RMS YSO sample are naturally clustered on this typical angular scale then the intrinsic clustering may mimic the apparent overdensity of YSOs observed in the surface density histogram and angular cross-correlation plot (Figs.~\ref{fig:surfdens} \& \ref{fig:crosscol}).

To compare the clustering of the control sample to that of the RMS YSO sample associated with bubbles we calculate the autocorrelation of the sample as a function of angular distance, using the estimator of \citet{landy1993} i.e.

\begin{equation}
\omega(\theta) = \frac{N_{DD} - 2N_{DR}  + N_{RR}}{N_{RR}}
\end{equation}

$N_{DD}$, $N_{DR}$ and $N_{RR}$ represent the normalised number counts of data-data, data-random and random-random pairs respectively. We calculate the autocorrelation for three separate samples: the entire RMS YSO sample, those associated with \emph{Spitzer} bubbles (i.e. lying within 2 bubble radii of a particular bubble), and those not associated with any \emph{Spitzer} bubbles (i.e. lying more than three bubble radii from \emph{all} bubbles). For simplicity we respectively refer to these samples as the full RMS YSO sample, the bubble-associated YSO sample and the control sample. 

We chose a value of 2 bubble radii for the radius of association  due to the steep fall off in surface density and angular cross-correlation beyond this radius. A total of 116 RMS YSOs are found within 2 bubble radii of a \emph{Spitzer} bubble and 629 RMS YSOs are found at an angular distance greater than 3 bubble radii from any bubble.  As in Sect.~\ref{sect:crosscol}, we constructed random catalogues of each of the three samples and performed 50 realisations of each random catalogue to avoid introducing higher levels of noise. 

\begin{figure}
\includegraphics[width=84mm,angle=-90]{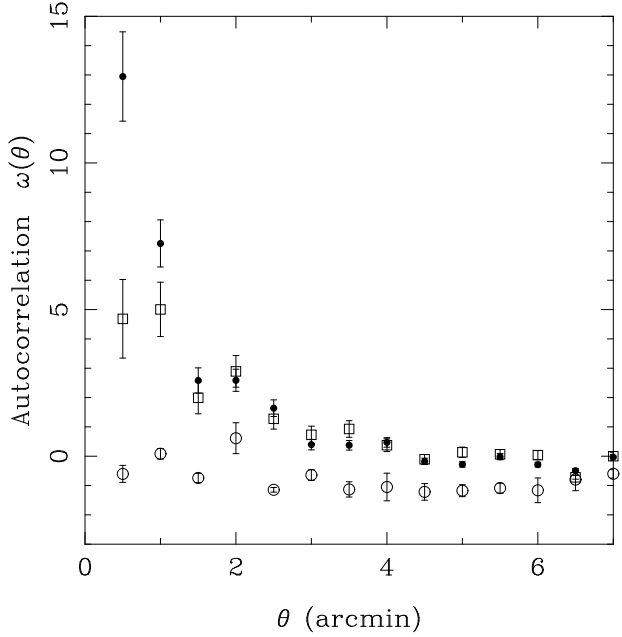}
\caption{The angular autocorrelation function of RMS YSOs for the full sample of RMS YSOs (solid dots), a ``control'' sample of RMS YSOs that lie more than 3 bubble radii from all of the \citet{churchwell2006} bubbles (open squares), and the sample of RMS YSOs that lie within 2 bubble radii of the \citet{churchwell2006} bubbles(open circles). 1$\sigma$ error bars are shown and are calculated using the bootstrap replacement method.}
\label{fig:autocol}
\end{figure}

The autocorrelations for these three samples of RMS YSOs are shown in Fig.~\ref{fig:autocol}. The behaviour of the full RMS YSO sample (solid dots in Fig.~\ref{fig:autocol}) shows a classic peak towards smaller angular scales and a decrease towards larger angular scales. This implies that the full RMS YSO sample is strongly clustered on scales of $\sim$1\arcmin\ or less. The bubble-associated YSO sample displays a markedly different behaviour, being \emph{anticorrelated} on all angular scales except for a small positive correlation at 2\arcmin . Finally, the control  sample shows a correlation on small angular scales similar to the full sample, although much weaker and with a flatter fall-off to large angular scales than the full RMS YSO sample. Interestingly there is a minor peak in the autocorrelation function at 2\arcmin, though this is not a statistically significant detection. At angular scales $\ge$2\arcmin\ the autocorrelation functions of the full RMS YSO sample and the control sample  are identical.

Clearly the autocorrelation function of the bubble-associated RMS YSOs is very different to the other two samples. This implies that the peaks in the YSO surface density and the YSO-bubble angular cross-correlation seen towards the bubble rims are not due to intrinsic clustering within the full RMS YSO sample. The YSOs in the full and control samples are much more highly correlated (i.e.~clustered) on smaller angular scales than the bubble-associated YSO sample. On the majority of angles the bubble-associated sample are anticorrelated, which implies that there is a shortfall in the numbers of YSOs associated with the bubbles at these angular separations compared to a random sample. The exception to this is at an angular scale of 2\arcmin, which corresponds closely with the median bubble diameter of  2.2\arcmin. This positive correlation suggests that we may be seeing the signature of RMS YSOs located on either side of the bubble, and the anticorrelation implies that the YSOs are not found on angular scales smaller or larger than this.

We do see a minor secondary peak at 2\arcmin\ in the autocorrelation functions of the full and control YSO samples, which at first sight suggests that a fraction of these YSOs are correlated at the same angular scale as the median bubble diameter. We expect this behaviour in the full YSO sample (which obviously includes YSOs associated with the bubbles). However, due to the incompleteness of the \citet{churchwell2006} catalogue caused by the manual search procedure that was used in its construction, we cannot exclude the presence of contaminating bubbles in our control sample.
Thus the autocorrelations that we measure may be artificially enhanced  for the control sample,  for example at the secondary peak at 2\arcmin. However, this peak is not statistically significant and  the full and control samples are clearly more strongly correlated at small scales than the bubble-associated sample. This suggests that the intrinsic clustering of RMS YSOs does not produce the enhancement of YSOs projected against the rims of the \emph{Spitzer} bubbles. Further work on the much more complete Milky Way Project bubble sample (Simpson et al.~2012, in prep) would aid this analysis.

\section{Discussion}
\label{sect:discussion}

In the following we combine our results on  the YSO surface density distributions, the YSO-bubble angular cross-correlation and YSO angular autocorrelation together with the measured properties of the bubbles and RMS YSOs. Our particular aims are to investigate the star formation environment of the bubbles in order to determine whether the YSO population is significantly enhanced near the bubbles and if there are any discernible differences in the population of YSOs found near the bubbles when compared to the entire sample. Ultimately we would like to place statistical constraints on the star formation associated with the bubbles that can inform current and future models of triggered star formation.

\subsection{The properties of \emph{Spitzer} bubbles associated with RMS YSOs.}
\label{sect:bubbleprops}

Taking the radius of association between YSO and bubbles to be 2 bubble radii  we find that a total of 116 YSOs and UC HIIs from the RMS YSO sample are associated with bubbles from the \citet{churchwell2006} catalogue. The number of bubbles from the \citet{churchwell2006} catalogue associated with one or more RMS YSOs is 72, which corresponds to 22 $\pm$ 3\% of the bubble catalogue (where the error is calculated using Poisson statistics). Of the 72 bubbles associated with RMS YSOs, 30 bubbles are associated with more than one RMS YSO or UC HII, ranging from 2 to 5 objects per bubble (and with a mean of 2.6 objects per bubble for the multiple matches). 

The majority of the \citet{churchwell2006} bubbles are not associated with RMS YSOs. This does not imply that the remaining bubbles are devoid of surrounding star formation, but merely that any star formation which is present is of sufficiently low luminosity to fall below the RMS detection threshold (which is typically $\ge$1000 L$_{\odot}$ at distances of a few kpc, \citealt{urquhart2011}). 
Unfortunately the majority of the \citet{churchwell2006} bubbles do not have measured distances that are free from  the kinematic distance ambiguity \citep{deharveng2010} and so we cannot determine the individual completeness limit for each bubble. However, the typical distances of the bubbles range from 2--13 kpc \citep{deharveng2010} and so we can confidently say that  the present RMS study of bubbles is complete to YSOs with a luminosity $\ge10^{4}$ L$_{\odot}$ for the most distant bubbles in the sample \citep{urquhart2011} and to YSOs with luminosity $\ge1000$ L$_{\odot}$ for bubbles located at the typical distance of a few kpc. 

As the luminosity of a B3 star is $\sim$550 L$_{\odot}$ \citep{meynet2000} we thus identify \emph{massive} star formation associated with the bubbles. Hence,  22 $\pm$ 3\% of the bubbles are associated with massive star formation in the MYSO or UC HII region phase. This is a similar fraction to that found by \citet{deharveng2010} and \citet{watson2010} who respectively found 18\% and 20\% of their bubble samples to be associated with either ultracompact HII regions, YSOs or 6.7 GHz methanol masers. The remaining 78 $\pm$ 3\% of the bubbles may be associated with low to intermediate mass star formation, but without a more sensitive survey we cannot confirm this hypothesis. The currently underway \emph{Herschel} Hi-GAL survey of the Galactic Plane \citep{molinari2010b,molinari2010a}  will provide such a sensitive survey of the entire bubble sample and it would be advantageous to revisit the bubbles with the more sensitive Hi-GAL data when it is available.

We searched for differences in the properties of the bubbles that are associated with RMS YSOs and those that are not, in order to try and identify differences in the properties of bubbles that are associated with massive star formation and those that are not. The two most important measured properties of each bubble are the size (strictly the mean radius of the bubbles) and thickness of the diffuse mid-infrared emission comprising the bubble. For both of these properties we determined the mean values for the 22\% of bubbles that are associated with at least one RMS YSO (i.e. within 2 bubble radii) and the remaining 78\% of bubbles that are not associated with an RMS YSO.


The mean radius of bubbles that are associated with an RMS YSO is 3.6\arcmin$\pm$0.4\arcmin, compared to the mean radius of unassociated bubbles of 4.6\arcmin$\pm$0.3\arcmin. The mean thickness of bubbles that are associated with RMS YSOs is 0.92\arcmin$\pm$0.08\arcmin, again compared to the unassociated bubble thickness which is 1.18\arcmin$\pm$0.07\arcmin. Quoted errors are the standard error on the mean. In order to determine the significance of these differences in mean radius and thickness we performed a two sample unequal variance (heteroscedastic) t test on each pair of samples (RMS YSO-associated bubbles and unassociated bubbles). The t tests return probabilities of  1\% and 0.8\% respectively that the mean radius and thickness of YSO-associated bubbles and unassociated bubbles are  drawn from populations with the same mean.

Those bubbles that are associated with massive star formation thus tend to be both smaller and thinner than those bubbles that are not associated with massive star formation. These results are not significant at a level of 3$\sigma$ or greater and, combined with the potential biases in the bubble population discussed earlier,  should be interpreted with caution. However, it is instructive to speculate on what may be the physical causes behind these observed differences.   \citet{weaver1977} predict that the radius of a wind-blown bubble should increase much faster than the thickness of the swept up shell when the bubbles are in their first expansion stage. Hence, bubbles with small radii and thinner shells should be younger than larger bubbles with thicker shells, as also suggested by \citet{dale2009}. Of course, we do not have physical distances for the majority of our sample and are thus dealing with angular radii rather than physical radii. Hence the bubbles that we have identified as small in angular size may just be the more distant members of the sample. However, over the whole sample of bubbles these effects should average out and our tentative results suggest that it is the younger bubbles within the sample that are more likely to be associated with massive star formation.

Theoretical models of shell fragmentation \cite[the collect and collapse process:][]{whitworth1994, dale2007} suggest that fragmentation of the swept up shell of a bubble tends to occur on timescales of one to a few Myr. Very little detailed study of the bubble lifetimes has currently been made. The few bubbles that have been studied to date have dynamical lifetimes of 0.5 to a few Myr \citep{watson2008,watson2009} and many of the sample are likely to be HII regions powered by late O to early B stars \citep{deharveng2010,bania2010,beaumont2010,anderson2011}, which again have main sequence lifetimes on the order of a few Myr to a few tens of  Myr . So the picture that massive star formation tends to be associated with smaller and younger bubbles is largely consistent with the predictions of the collect \& collapse models that star formation should happen on $\sim$ Myr timescales. However, if these bubbles are shown to be much younger than 0.5 Myr then this would point towards the star formation being caused by the implosion of pre-existing density structures \citep[e.g.~the Enhancement of initial Density substructure and simultaneous Global Implosion proposed by][]{walch2011}, or perhaps simply pre-existing untriggered star formation as suggested by \citet{dale2011}.

In the following  subsection we concentrate upon the surface density of YSOs around the bubbles, with the aim of showing that there is a statistically significant overdensity of YSOs and that it is unlikely that the majority of the YSOs formed spontaneously.

\subsection{An overdensity of YSOs around \emph{Spitzer} bubbles}
\label{sect:overdensity}

It is clear from the surface density plots shown in Figures \ref{fig:surfdens}--\ref{fig:mmb_surfdens} that there is a significantly enhanced surface density of YSOs found at the rim of the bubbles. For the RMS YSO sample shown in Fig.~\ref{fig:surfdens} the YSO surface density is 19.1$\pm$ 3.6 YSOs/unit area at an angular distance from the bubble centre of  1 bubble radius, some 4$\sigma$ above the mean surface density of YSOs at angular distances greater than 2 bubble radii (3.1$\pm$0.2 YSOs/unit area). This result is confirmed by the angular cross-correlation function of the RMS YSO sample shown in Fig.~\ref{fig:crosscol} where the cross-correlation function peaks at an angular distance of 1 bubble radius with a significance of 9$\sigma$. In the immediate environment of a \emph{Spitzer} bubble the highest probability location to find an RMS YSO is projected against the rim of the bubble.

Moreover, it is clear from inspecting Fig.~\ref{fig:surfdens} that the surface density of YSOs is not only enhanced at an angular offset of 1 bubble radius, but that it is enhanced over the entire angular scale of the bubbles out to an angular offset of 2 bubble radii.  We can see this by comparing the mean surface density of YSOs ``inside'' 2 bubble radii and ``outside'' 2 bubble radii. The mean surface density of YSOs within an angular offset of 2 bubble radii is 8.9$\pm$1.7 YSOs/unit area compared a value of 3.2$\pm$0.2 YSOs/unit area at an angular offset of 2 bubble radii or greater. A two sample unequal variance (heteroscedastic) t test of these two subsamples returns a probability of only 0.4\% that these two subsamples are drawn from populations with the same mean. Hence we have demonstrated that there is a statistically significant overdensity of massive YSOs associated with the bubbles compared to the background, with an enhanced probability of finding these YSOs projected against the rim of the bubbles.


What do these results imply? Firstly, there is a greater concentration of massive star formation towards the bubbles than in the wider environment. This result is  confirmed by the surface density of MMB 6.7 GHz masers (see Fig.~\ref{fig:mmb_surfdens}), which trace a YSO population independently of mid-infrared emission. A greater concentration of star formation towards the bubbles implies that the bubbles are either efficient at producing YSOs, or that they are found in regions of high YSO surface density.  This is the classic chicken and egg scenario applied to massive star formation: do the bubbles precede the high surface density of YSOs, or does the high surface density of YSOs precede (or occur simultaneously with) the formation of the bubbles?

Before considering this question more fully, we must bring in the second of our results -- that there is an enhanced probability of finding YSOs projected against the rim of the bubbles (i.e. at an angular offset of 1 bubble radius). By inspecting the autocorrelation of the RMS YSOs we showed in Section \ref{sect:autocol} that this effect is not likely to be due to intrinsic clustering within the RMS sample on similar angular scales to the bubble radii. The ancillary question raised by this result is: why are the YSOs more likely to be found projected against the rim of the bubbles. i.e. what is special about the bubble rims? 

The bubble rims are traced by 8$\mu$m PAH emission which originates from the photon-dominated region between the ionisation front being driven out by the HII region within the bubble and the surrounding neutral medium. The rim of the bubbles thus shows the interface between HII region and surrounding neutral gas. For a spherical bubble morphology one would expect the column density of gas to be greater at the bubble rims due to the greater path length through the neutral material towards the rims. So at first glance, the higher surface density of YSOs projected against the bubble rims may simply reflect the higher column density at the rim of the bubbles, i.e. the YSOs trace molecular column density.

However, while the sample of bubbles that have been observed at relatively high angular resolution in molecular lines \citep{beaumont2010} do show a peaked molecular column density profile at a normalised bubble radius of 1, the column density falls off much less sharply than the YSO surface density. Inspection of Figure 2 from \citet{beaumont2010} shows that at a normalised bubble radius of 1.5 the CO intensity can be roughly half of that at a normalised radius of 1. This suggests that the YSOs may not  trace the column density distribution, although much closer scrutiny of the bubbles in a non-optically thick tracer is required to confirm this hypothesis. Moreover, the CO contrast between the centre of the bubbles and their rims is often extreme \citep{beaumont2010} whereas the YSO surface density within an angular offset of 2 bubble radii is everywhere higher than the background level. Thus we cannot confidently say that the YSO surface density traces the gas column density around the bubbles.

The YSO surface density is strongly peaked at an offset of 1 bubble radius and decreases sharply beyond this value. Beyond an angular offset of 2 bubble radii the surface density of YSOs is essentially undistinguishable from the background level. The angular cross-correlation function shows a similar  steep drop --- beyond an angular distance of 2 bubble radii the bubbles and RMS YSOs are essentially \emph{uncorrelated}. The implication of this is that whatever causes the rise in YSO surface density is closely related to the rim of the bubbles. The bubble radius is a dynamic value and expected to increase over time as stellar winds or radiation pressure causes the bubbles to expand. Combined with this is the fact that the massive YSOs and UC HII regions identified by the RMS survey typically tend to have lifetimes around  a few 10$^{4}$ to a few $10^{5}$ years \citep{mottram2011b} and so should trace very recent star formation. 

The sum of these pieces of information leads us to conclude that it is likely that the bubbles predate the YSOs.  If the bubbles formed in an environment with a high surface density of YSOs \citep[e.g.~in the turbulent highly fragmented initial conditions suggested by][]{dale2011}, then the distribution should not peak at the rim of the bubble as the bubble radius is time-dependent. Similar arguments have been used by \citet{preibisch2011} for  YSOs detected at the edges of shells  in Carina. Also in this case the  extent of the enhanced YSO surface density should also not be related to the current radius of the bubble -- why are bubbles found in regions of enhanced YSO surface density occupying twice their \emph{current} angular radius? Finally, the relative timescales of the massive YSOs and those required for the expansion of the bubbles imply that the YSOs formed \emph{after} the bubbles. We thus conclude that a significant fraction of the YSOs seen against the rim of the bubbles were likely triggered by the expansion of the bubble.

A greater understanding of the dynamical timescales for the expansion of the bubbles and also the molecular environment of the bubbles are required to confirm this hypothesis. Pinpointing the YSO formation to have occurred after the bubble was formed is crucial to disentangling cause and effect in the star formation surrounding the bubbles. Currently, only a few bubbles have had their dynamical lifetimes estimated and more studies similar to those of \citet{watson2009} are required over a larger sample of bubbles. Comparing the YSO distribution to the gas distribution is also crucial to investigate differences in the population of YSOs at the rims of \emph{Spitzer} bubbles, for example to determine whether the star formation efficiency is  enhanced at the bubble rims. Finally, it will also be instructive to apply the same statistical tools that we have used in this paper to the latest generation of triggered star formation models \citep[e.g.][]{walch2011,dale2011} in order to see whether the surface density of sink particles in the SPH simulations matches that of YSOs around observed bubbles.

\subsection{The luminosity function of RMS YSOs associated with bubbles}

\begin{figure}
\includegraphics[width=84mm]{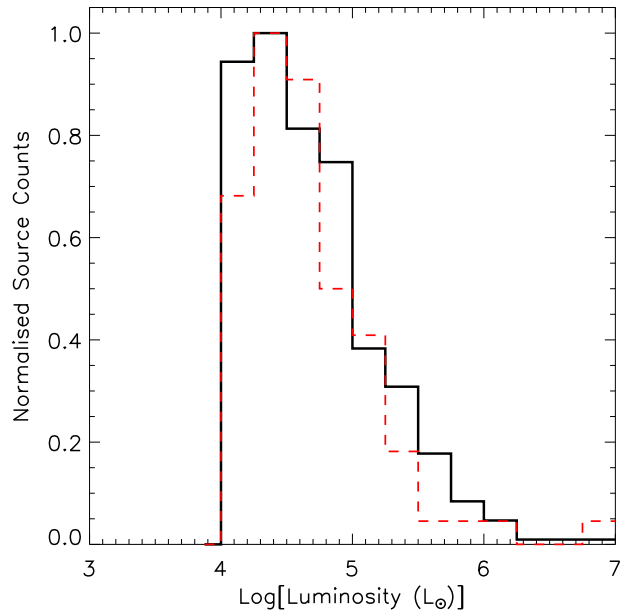}
\caption{The luminosity function of the entire RMS YSO sample (solid line) and those YSOs that lie within an angular offset of 2 bubble radii from a \emph{Spitzer} bubble (dashed line). }
\label{fig:lumfunc}
\end{figure}

\begin{figure}
\includegraphics[width=84mm]{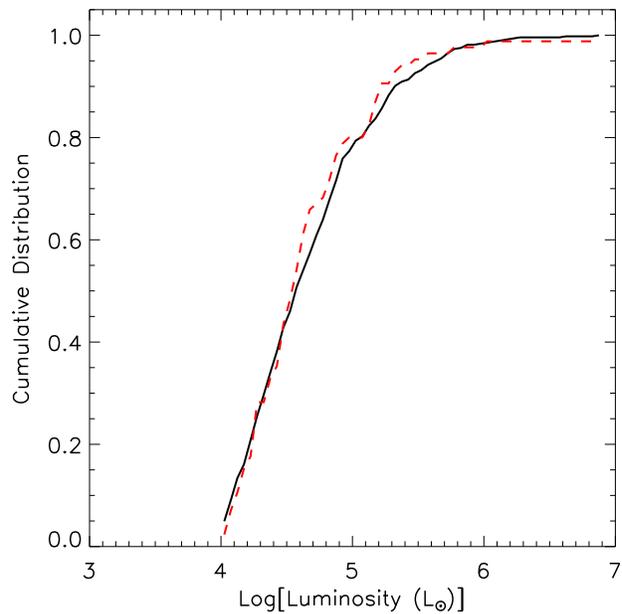}
\caption{The cumulative distribution function of the  luminosities  of the entire RMS YSO sample (solid line) and those YSOs that lie within an angular offset of 2 bubble radii from a \emph{Spitzer} bubble (dashed line). }
\label{fig:lumfunc_cdf}
\end{figure}

The RMS survey has determined luminosities for their entire sample of YSOs and UC HII regions \citep{mottram2011a,urquhart2011}, and so in this subsection we seek to identify differences in the luminosity function between  those YSOs that are associated with bubbles (i.e. within an angular offset of 2 bubble radii) compared to the full population. As both luminosity functions are essentially power laws with turnovers caused by incompleteness at low luminosities, one must take care that any differences between the two samples are not primarily due to differences in the completeness of each sample. To avoid this issue we cut both samples at a luminosity of 10$^{4}$ L$_{\odot}$, i.e. the minimum YSO luminosity at which the RMS catalogue is complete \citep{urquhart2011}.
We plot these truncated luminosity functions in Fig.~\ref{fig:lumfunc} and the corresponding cumulative distribution functions in Fig.~\ref{fig:lumfunc_cdf}.

Inspecting Fig.~\ref{fig:lumfunc} shows that the RMS YSOs associated with the bubbles do not tend to have a higher luminosity than the rest of the population. The two luminosity functions are essentially identical. A Kolmogorov-Smirnov test on the two luminosity functions yields the result that there is a 27\% probability that the two distributions are drawn from the same sample, and thus we cannot reject the null hypothesis with any significance. If the YSOs associated with the bubbles were triggered (as we argue in Sect.~\ref{sect:overdensity}) then we conclude that the triggering process does not result in stars with an appreciably different luminosity function.

This result must be contrasted against the many inferences that have been made to date which suggest that the triggering process may produce stars with a luminosity function skewed to higher luminosities \citep[e.g.][]{sugitani1989,sugitani1991,dobashi2001, deharveng2005, motoyama2007,urquhart2009c}. Fragmentation models for swept-up shells around massive stars predict a top-heavy mass distribution of fragments, which in turn may lead to a top-heavy IMF distribution \citep{whitworth1994,dale2009}. Observations of molecular clouds associated with HII regions also suggested that IRAS point sources associated with the clouds nearer the HII regions are more luminous than those found to be more distant \citep{yamaguchi1999}. However these results are not consistent with the largely constant form of the IMF observed over Galactic scales. If the triggering process does induce a top-heavy IMF then the results of triggering do not dominate the IMF \citep{dale2009}. Here we have shown that if these YSOs are triggered, then the triggering process results in a luminosity function that is indistinguishable from the full YSO sample.

\subsection{The fraction of massive stars in the Milky Way that may have been triggered}

Estimating the impact of the triggering process in global Galactic star formation  is a vital part of determining a simple prescription for star formation that can be applied to galaxy evolution models. Because we have combined the results of two large area and relatively unbiased surveys (the RMS survey and the \citet{churchwell2006} bubble catalogue) we are in a position to try and estimate the contribution of triggered star formation to the Galaxy's population of stars. Before doing so, we must stress the major caveat involved -- that the \citet{churchwell2006} bubble catalogue is likely to be highly incomplete. Thus, we do not attempt a detailed treatment at this stage and simply infer a lower limit to the fraction of massive stars in the Milky Way that \emph{may} have been triggered.  

As discussed in Sect.~\ref{sect:bubbleprops} we find 116 YSOs and UC HIIs from the RMS YSO sample that are associated with bubbles from the \citet{churchwell2006} catalogue.  The bubble-associated YSOs form 14\% of the 846 objects contained in the RMS YSO sample over the GLIMPSE survey region. The RMS survey is complete for essentially all massive YSOs with luminosities in excess of $10^{4}$ L$_{\odot}$ out to the furthest kinematic distance in the bubble sample ($\sim$14 kpc). Thus, assuming that the incompleteness of the Churchwell catalogue dominates over the fraction of the bubble-associated YSOs that were triggered, we estimate that at least 14\% of the massive stars in the Miilky Way could have been triggered. If, as suggested by \citet{churchwell2006}, the \emph{Spitzer} bubble catalogue is $\sim$50\% incomplete then the true fraction of triggered massive stars could be up to $\sim$30\%.

\section{Summary and Conclusions}
\label{sect:conclusions}

We have carried out a detailed statistical study of young stellar objects (YSOs) found nearby \emph{Spitzer} bubbles from the \citet{churchwell2006} catalogue. Our main conclusions are summarised below:

\begin{enumerate}

\item The surface densities of YSOs and UC HIIs from the RMS survey \citep{urquhart2011}, intrinsically red sources from \citet{robitaille2008} and MMB 6.7 GHz methanol masers \citep{green2009} are enhanced towards \emph{Spitzer} bubbles from the \cite{churchwell2006} catalogue. The surface density of all three YSO catalogues peaks toward the projected angular radius of the bubbles, with a peak surface density of 4$\sigma$ above the mean background level for RMS YSOs and UC HII regions. The mean surface density of YSOs associated with the bubbles is  overdense with respect to the surrounding mean background at the 3$\sigma$ level.

\item The angular cross-correlation function of RMS YSOs and \emph{Spitzer} bubbles shows a similar behaviour to the surface density, with a  9$\sigma$ peak in the cross-correlation function at an angular offset of 1 bubble radius. In the immediate environment of a \emph{Spitzer} bubble the highest probability location to find an RMS YSO is projected against the rim of the bubble. RMS YSOs and \emph{Spitzer} bubbles are essentially uncorrelated beyond an angular distance of 2 bubble radii. Examination of the autocorrelation functions suggests that these effects are not caused by intrinsic clustering within the RMS YSO sample.

\item Most \emph{Spitzer} bubbles are not associated with massive YSOs: 22 $\pm$ 3\% of the Churchwell catalogue are associated with RMS YSOs. This fraction is consistent with smaller studies \citep{deharveng2010,watson2010}.

\item \emph{Spitzer} bubbles associated with RMS YSOs tend to possess both thinner rims and smaller angular radii than bubbles that are not associated with RMS YSOs. We interpret this tendency to be due to an age effect, with RMS YSOs forming around younger and smaller bubbles.

\item The different relative timescales for the formation of \emph{Spitzer} bubbles and  RMS YSOs, and the strong peak in surface density and cross-correlation at the rims of the bubbles lead us to conclude that a significant fraction of the RMS YSOs were triggered by the expansion of the bubble.

\item We find no significant  differences in the luminosity  function of RMS YSOs associated with \emph{Spitzer} bubbles compared to the entire RMS YSO population, which suggests that the triggering process does not result in a top heavy luminosity function or IMF.

\item We estimate from the fraction of RMS YSOs associated with bubbles and the incompleteness of the bubble catalogue that the lower limit for the fraction of massive stars in the Milky Way that could have been triggered is 14\%. If the \citet{churchwell2006} catalogue is 50\% incomplete then the upper limit to the fraction of massive stars that could have been triggered may be up to $\sim$30\%. Therefore this mode of massive star formation ought not to be ignored when considering star formation on Galactic scales.

\end{enumerate}

We must stress that these results are based on the current and largely incomplete sample of known \emph{Spitzer} bubbles, which were identified by manual searches of the GLIMPSE image database \citep{churchwell2006}. However the groundwork for a much more comprehensive catalogue of bubbles is currently being laid by the Milky Way Project\footnote{\texttt{http://www.milkywayproject.org}}, which aims to identify many more bubbles by a systematic citizen science survey of the GLIMPSE images (Simpson et al 2012, in prep). In addition, the \emph{Herschel} Hi-GAL survey \citep{molinari2010a} holds the promise of a much more complete YSO catalogue reaching fainter luminosities than the RMS Survey.

Statistical studies of the type that we have presented here will be of increasing importance in the age of large scale surveys of the Milky Way, and offer the prospect of being able to directly compare the observed distribution of YSOs to the predicted distribution of sink particles in triggered star formation models \citep[e.g.][]{walch2011,dale2011}. With these advances the study of triggered star formation  will finally move beyond the phenomenological stage to be able to make direct predictions.

\section*{Acknowledgments}

We would like to thank the anonymous refree for a report that was both prompt and illuminating. We thank Steve Longmore and Willem Baan for useful and illuminating discussions on potential biases in YSO identification and the incompleteness of the bubble catalogue. We would also like to thank Andrew Walsh for organising a stimulating meeting of observers and theorists in Townsville that provided the initial seed for this work. This research would not have been possible without the  NASA Astrophysics Data System Bibliographic 
Services. This paper made use of information from the Red MSX Source survey database at www.ast.leeds.ac.uk/RMS which was constructed with support from the Science and Technology Facilities Council of the UK. We made use of positional data on 6.7 GHz methanol masers provided by  the Methanol MultiBeam (MMB) Survey, also supported by the Science and Technology Facilities Council, and with a maser database available at \texttt{http://astromasers.org}

\bibliography{./standardbibtex}
\bibliographystyle{mn2e}

\bsp

\label{lastpage}

\end{document}